\documentclass[journal]{IEEEtran}

\ifCLASSINFOpdf
\else
   \usepackage[dvips]{graphicx}
\fi
\usepackage{url}

\usepackage{multirow}

\usepackage{booktabs}
\usepackage{cite}
\usepackage{algorithm}
\usepackage{algorithmicx}
\usepackage{algpseudocode}
\usepackage{comment}
\usepackage{subcaption}

\newcommand{\Input}{\item[\textbf{Input:}]}
\newcommand{\Output}{\item[\textbf{Output:}]}

\usepackage{textcomp}
\usepackage{graphicx}
\usepackage{amsmath,amssymb,amsfonts}
\usepackage{acronym}
\newacro{ASD}[ASD]{Anomalous Sound Detection}
\newacro{LDM}[LDM]{Latent Diffusion Model}
\newacro{GANs}[GANs]{Generative Adversarial Networks}
\begin{document}

\title{ESTM: An Enhanced Dual-Branch Spectral-Temporal Mamba for Anomalous Sound Detection}

\author{Chengyuan Ma, Peng Jia, Hongyue Guo, and  Wenming Yang, \IEEEmembership{Senior Member, IEEE}
\thanks{This work was supported in part by the National Key R\&D Program of China (2023YFB4302200).}
\thanks{Chengyuan Ma and Wenming Yang are with the Shenzhen International Graduate School, Tsinghua University, Shenzhen 518055, China (Email: mcy23@mails.tsinghua.edu.cn; yang.wenming@sz.tsinghua.edu.cn), Wenming Yang is the corresponding author.}
\thanks{Peng Jia and Hongyue Guo are with the Collaborative Innovation Center for Transport Studies, Dalian Maritime University, Dalian 116026, China.(
Email: jiapeng@dlmu.edu.cn; hyguo@dlmu.edu.cn)}
}

\maketitle

\begin{abstract}
The core challenge in industrial equipment anomalous sound detection (ASD) lies in modeling the time-frequency coupling characteristics of acoustic features. Existing modeling methods are limited by local receptive fields, making it difficult to capture long-range temporal patterns and cross-band dynamic coupling effects in machine acoustic features. In this paper, we propose a novel framework, ESTM, which is based on a dual-path Mamba architecture with time-frequency decoupled modeling and utilizes Selective State-Space Models (SSM) for long-range sequence modeling. ESTM extracts rich feature representations from different time segments and frequency bands by fusing enhanced Mel spectrograms and raw audio features, while further improving sensitivity to anomalous patterns through the TriStat-Gating (TSG) module. Our experiments demonstrate that ESTM improves anomalous detection performance on the DCASE 2020 Task 2 dataset, further validating the effectiveness of the proposed method.

\end{abstract}


\IEEEpeerreviewmaketitle

\vspace{-2mm}
\section{Introduction}
\begin{figure*}[h]
    \centering
    \includegraphics[width=0.8\textwidth]{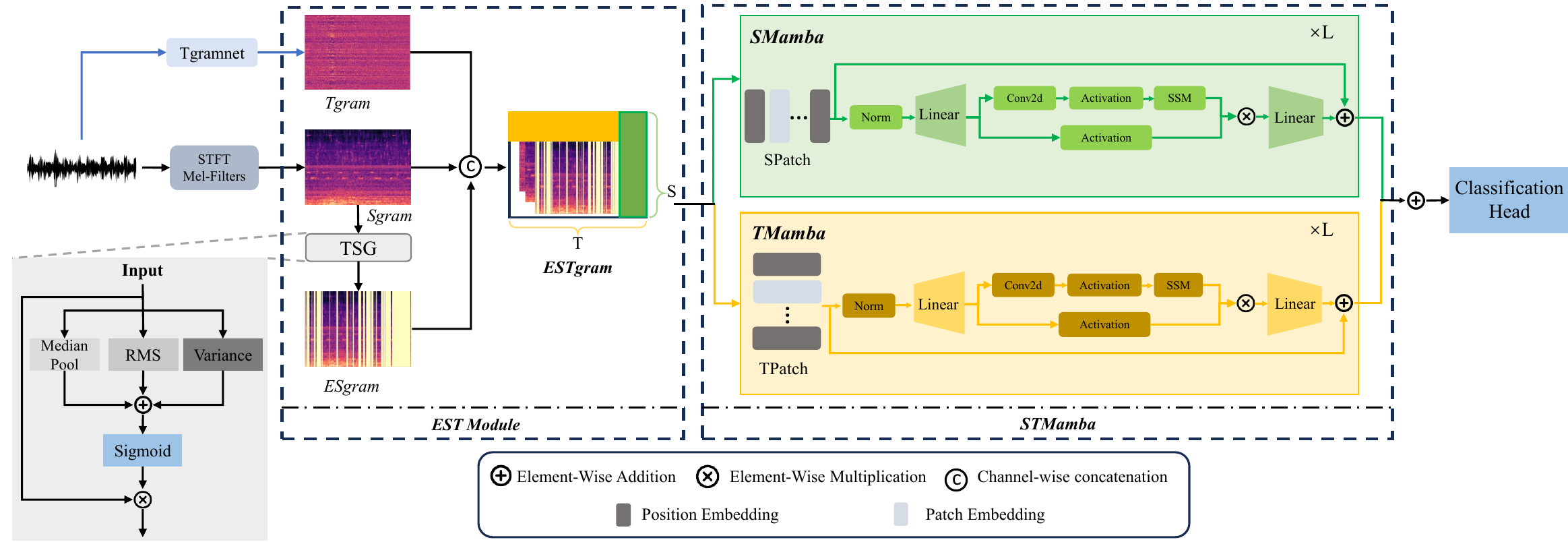}
    \caption{
    Framework of the Enhanced  Spectral Temporal Mamba(ESTM).
    }
    \label{fig:1}
\end{figure*}

\IEEEPARstart{W}{ith} the widespread application of audio analysis in machine condition monitoring, anomalous sound detection has attracted significant attention from researchers. Anomalous Sound Detection (ASD) has become a research focus due to its diagnostic capability in unsupervised scenarios\cite{Koizumi_DCASE2020_01,suefusa2020anomalous,giri2020self}. This task requires constructing a generalized detection ability for unknown anomalous patterns under training conditions that include only normal samples, which fundamentally differs from traditional supervised acoustic classification. Existing methods typically distinguish between normal and anomalous sounds by learning the feature distribution of normal sounds.

Current methods can be categorized into two main types, both of which have corresponding limitations. Reconstruction error-based methods (e.g., ID-conditioned AE \cite{kapka2020id}, Glow-Aff \cite{dohi2021flow}) suffer from parameter redundancy and lack cross-machine generalization capability due to device-specific modeling. Another approach applies metadata (machine type and machine ID) as labels to classify audio features and computes anomaly scores based on classifier output probabilities. These methods \cite{liu2022anomalous,chen2023sw,25icassp,Noisy-Arcmix,first-shot} have achieved state-of-the-art (SOTA) performance in ASD and require training only a single model. Although metadata-driven approaches provide a unified architecture, their frame-level processing mechanism introduces two main issues: (i) local time windows struggle to capture the long-period steady-state characteristics of machine operations; (ii) independent frame-wise spectral modeling ignores the coordinated variation of energy distribution across frequency bands. Some studies \cite{zhang2024dual,guan2023time,cst} attempt to capture salient patterns in both the frequency and time domains to enrich learned representations, but they still fail to capture trend-level anomalous patterns effectively.

Recent advancements in Selective State-Space Models (SSMs) \cite{gu2023mamba,10.5555/3692070.3692469,wang2023ssm} provide a new perspective to address these challenges. The Mamba architecture incorporates an input-dependent state transition mechanism, which enables it to adaptively focus on critical patterns within sequences, including audio features \cite{xiao2024tf,audio-mamba,ssamba,mamba-speech}. This characteristic is highly aligned with the requirements of the ASD task: 
\begin{itemize}
\item long-term contextual modeling of steady-state operational noise in machines; 
\item localized sensitivity regulation for sudden anomalous sound segments.
\end{itemize}
Based on this, we propose the Enhanced  Spectral Temporal Mamba framework (ESTM), introducing selective state-space modeling with decoupled time-frequency processing into the ASD field for the first time. The STMamba backbone extracts feature representations from frequency patches and time patches, which are segmented after fusing raw audio features and enhanced Log-Mel spectrograms. The model is trained in a classification manner using audio metadata as class labels. To the best of our knowledge,  ESTM is the first to introduce a decoupled dual-path Mamba architecture specifically tailored for the challenges of ASD. We evaluate our method on the DCASE 2020 Task 2 dataset, demonstrating its effectiveness across various machine types. Furthermore, ablation experiments validate the effectiveness of each designed module.

The main contributions of this paper are as follows:
\begin{itemize}
\item We propose ESTM, a dual-path Mamba architecture designed with both frequency-guided and time-guided branches. It utilizes learnable spectral-temporal patches to differentiate and enhance feature representations while employing selective state-space modeling for global sequence modeling over long time spans, enriching feature representations across different frequency bands and time periods and making the model more sensitive to various anomalous pattern changes.

\item We propose a TriStat-Gating (TSG) module, which introduces a novel input feature, ESgram, an enhanced Mel spectrogram. This module applies a set of parameter-free multi-angle augmentations to the Mel spectrogram, improving the system’s sensitivity to unknown anomalies.

\item We demonstrate the feasibility of ESTM and achieve SOTA performance on the DCASE 2020 Task 2 dataset.
\end{itemize}

\vspace{-3mm}
\section{Proposed Method}

The overall architecture of the proposed ESTM is illustrated in Fig. 1. The input features to the backbone consist of three components: the raw audio feature Tgram, the Log-Mel spectrogram feature Sgram, and the enhanced Mel spectrogram feature ESgram, which are fused to form ESTgram. The spectrogram is segmented along both the time axis and the frequency axis, enabling time-frequency decoupled modeling through the dual-path STMamba. Finally, the extracted features are fused and fed into a classification head for classification. This section provides a detailed explanation of the proposed method.

\vspace{-3mm}
\subsection{Enhanced Spectral Temporal Fusion Feature}

The input features to the STMamba backbone are obtained as follows:

For a given raw single-channel machine audio signal $\mathbf{x} \in \mathbb{R}^{1 \times L}$, its Log-Mel spectrogram is denoted as $\mathbf{x}_{\mathrm{Mel}} \in \mathbb{R}^{M \times N}$, where $M$ represents the number of Mel bins (frequency width of the spectrogram), and $N$ represents the number of time frames (spectral height). To enhance the Log-Mel spectrogram from different perspectives, we employ the TriStat-Gating (TSG) module. It constructs an adaptive gating mechanism based on three-order statistical metrics: median \cite{gholamalinezhad2020pooling}, root mean square \cite{zhang2019root} (RMS), and variance \cite{toda2005spectral}, which correspond to the stability, intensity consistency, and distribution uniformity of the machine acoustic characteristics, respectively.

For the transposed Log-Mel spectrogram $\mathbf{x}_{\mathrm{Mel}}^{\top} \in \mathbb{R}^{N \times M}$,  frame-wise statistical calculations are performed so that each time frame independently generates a gating coefficient. This results in a gating vector for the time frames:
\begin{equation}
    Pool(\mathbf{x}_{\mathrm{Mel}}^\top)=\sigma\left(\alpha\left[\mathrm{M}(\mathbf{x}_{\mathrm{Mel}}^\top)+\mathrm{R}(\mathbf{x}_{\mathrm{Mel}}^\top)+\mathrm{V}(\mathbf{x}_{\mathrm{Mel}}^\top)-\mu\right]\right)
\end{equation}

where $\mathrm{M}(\mathbf{x}_{\mathrm{Mel}}^\top)$, $\mathrm{R}(\mathbf{x}_{\mathrm{Mel}}^\top)$, and $\mathrm{V}(\mathbf{x}_{\mathrm{Mel}}^\top) \in \mathbb{R}^N$ represent the median, root mean square, and variance vectors computed along the frequency dimension for each time frame, respectively. $\mu$ is the mean used for de-centering the gating vector, $\alpha > 0$ is a scaling factor, and $\sigma(\cdot)$ is the sigmoid function applied element-wise to the vector of length $N$.

The resulting gating vector is then broadcast along the frequency dimension and element-wise multiplied with the original spectrogram to produce the enhanced spectrogram feature $\mathbf{x}_{\mathrm{EMel}}$:

\begin{equation}
    \mathbf{x}_{\mathrm{EMel}}=\mathrm{Pool}(\mathbf{x}_{\mathrm{Mel}}^{\top})\otimes\mathbf{x}_{\mathrm{Mel}}
\end{equation}
where $\otimes$ denotes element-wise multiplication.

In addition to these features, we also incorporate the Tgram from \cite{liu2022anomalous} to supplement the information from the raw audio signal. The final ESTgram is constructed by concatenating $\mathbf{x}_{\mathrm{Mel}}$, $\mathbf{x}_{\mathrm{EMel}}$, and $\mathbf{x}_{\mathrm{T}}$ along the channel dimension:
\begin{equation}
    \mathbf{x}_{EST}=Concat(\mathbf{x}_{Mel},\mathbf{x}_{EMel},\mathbf{x}_T)
\end{equation}
where $Concat(\cdot)$ represents the concatenation operation along the channel dimension. We employ a set of orthogonal statistical metrics (median, RMS, and variance) to characterize the signal's stability, energy, and spectral shape, yielding a non-redundant feature representation sensitive to anomalous time-frequency patterns.
\begin{table*}[t]
\caption{
Performance comparison of AUC (\%) and pAUC (\%) across different machine types on the test data of the development and additional datasets.}

    \begin{tabular}{c|cccccccccccc|cc}
        \toprule
        \multirow{2}{*}{Methods} & \multicolumn{2}{c}{Fan} & \multicolumn{2}{c}{Pump} & \multicolumn{2}{c}{Slider} & \multicolumn{2}{c}{Valve} & \multicolumn{2}{c}{ToyCar} & \multicolumn{2}{c}{ToyConveyor} & \multicolumn{2}{c}{Average} \\
        \cmidrule(lr){2-3} \cmidrule(lr){4-5} \cmidrule(lr){6-7} \cmidrule(lr){8-9} \cmidrule(lr){10-11} \cmidrule(lr){12-13} \cmidrule(lr){14-15} 
         & AUC & pAUC & AUC & pAUC & AUC & pAUC & AUC & pAUC & AUC & pAUC & AUC & pAUC & AUC & pAUC \\
        \midrule

        MobileNetV2 \cite{giri2020self} & 80.19 & 74.40 & 82.53 & 76.50 & 95.27 & 85.22 & 88.65 & 87.98 & 87.66 & 85.92 & 69.71 & 56.43 & 84.34 & 77.74 \\
        Glow\_Aff \cite{dohi2021flow} & 74.90 & 65.30 & 83.40 & 73.80 & 94.60 & 82.80 & 91.40 & 75.00 & 92.20 & 84.10 & 71.50 & 59.00 & 85.20 & 73.90 \\
        STgram-MFN  \cite{liu2022anomalous} & 94.04 & 88.97 & 91.94 & 81.75 & 99.55 & 97.61 & 99.64 & 98.44 & 94.44 & 87.68 & 74.57 & 63.60 & 92.36 & 86.34 \\
        CLP-SCF \cite{zhang2024dual} &96.98 & 93.23 & 94.97 & 87.39 & 99.57 & 97.73 & 99.89 & 99.51 & 95.85 & 90.19 & 75.21 & 62.79 & 93.75 & 88.48 \\
        ASD-AFPA \cite{zhang23fa_interspeech} & 97.55  & 93.48  & 94.46  & 86.76  & 99.69  & \textbf{98.40}  & 99.12  & 95.42  & 96.15  & 89.45  & 76.49  & 64.21  & 93.91  & 87.95\\
        TASTgram(NAMix) \cite{Noisy-Arcmix} & 98.10  & 94.97  & 94.51  & 85.22  & 99.49  & 97.30  & \textbf{99.95}  & 95.72  & 96.19  & 89.96  & 76.26  & 66.95  & 94.08  & 89.01\\
        \textbf{ESTM(ours)} & \textbf{98.85} & \textbf{95.72}    & \textbf{96.39}    &  \textbf{87.60}    & \textbf{99.74}          & 98.18       & 99.92 & \textbf{99.77} & \textbf{97.28}    & \textbf{91.71}    & \textbf{82.35}  & \textbf{69.96} & \textbf{95.76} & \textbf{90.49} \\
        \bottomrule
    \end{tabular}

\label{tab:tab2}
\end{table*}
\vspace{-3mm}

\subsection{Spectral Temporal Mamba}

We use STMamba to extract features from the ESTgram, denoted as $\mathbf{x}_{EST} \in \mathbb{R}^{T \times S}$. Due to the one-dimensional sequential modeling architecture of Mamba, to achieve time-frequency decoupled feature modeling, we segment and flatten the enhanced ESTgram along both the time axis and frequency axis, generating unified representations $\mathbf{p}$. Let the time window size be $K$ and the frequency window size be $H$. For the $i$-th time patch, the feature representation is $\mathbf{p}_t^{(i)}\in\mathbb{R}^{W\cdot S}$, and for the $j$-th frequency patch, the feature representation is $\mathbf{p}_s^{(j)}\in\mathbb{R}^{T\cdot H}$ Inspired by VIM \cite{pmlr-v235-zhu24f}, we append a learnable class token $c_{*}$ in both the time direction and frequency direction as a global feature aggregator. Let the number of patches along the time and frequency axes be $I$ and $J$ , respectively. The tokens are mapped into a unified $D$-dimensional space, and positional embeddings $E_{pos}^{I}\in\mathbb{R}^{(I+1)\times D}$ and $E_{pos}^{J}\in\mathbb{R}^{(J+1)\times D}$ are introduced to incorporate positional information. The overall feature vectors are expressed as:
\begin{equation}
    X_I*=[c_{*};\mathbf{p}^1_{*}W;\mathbf{p}^2_{*}W;...;\mathbf{p}^{I}_{*}W]+E_{pos}^{I}
\end{equation}
\begin{equation}
    X_J*=[c_{*};\mathbf{p}^1_{*}W;\mathbf{p}^2_{*}W;...;\mathbf{p}^{J}_{*}W]+E_{pos}^{J}
\end{equation}
where $W$ is the projection matrix, and $X_I*$ and $X_J*$ denote the input sequences for the time and frequency axes, respectively.

The STMamba backbone consists of two components: SMamba and TMamba. SMamba performs top-down causal scanning along the frequency axis, leveraging gated convolution kernels to accumulate energy across frequency bands, enabling the extraction of local spectral features and global frequency-domain variations. TMamba scans left-to-right along the time axis, utilizing a delayed state cache to capture long-term temporal dependencies. Both branches are built using Selective State-Space Models (SSMs). The specific mathematical formulation of the modules can be expressed as:

\begin{algorithm}
\caption{Dual-Path Mamba Architecture}
\label{alg:bidirectional_mamba_block}
\begin{algorithmic}[1]
\small  
\Input Dual input embeddings: 
    \Statex $X_I^* = [c_{*};p^1_{*}W;\ldots;p^{I}_{*}W] + E_{pos}^{I}$ (Path-I, length $I$)
    \Statex $X_J^* = [c_{*};p^1_{*}W;\ldots;p^{J}_{*}W] + E_{pos}^{J}$ (Path-J, length $J$)
\Output Fused embeddings: $H$
\For{each input: $X \in \{X_I^*, X_J^*\}$}
    \State $X' \gets \text{LayerNorm}(X)$
    \State $X' \gets \text{Linear}(X')$ 
    \State $X' \gets \text{Conv2D}(X')$ 
    \State $X' \gets \text{Activation}(X')$
    \State $\Delta \gets \log(1 + \exp(\text{Linear}_\Delta(X')))$ 
    \State $A \gets \Delta \odot \text{Parameter}_A$, $B \gets \Delta \odot \text{Linear}_B(X')$
    \State $C \gets \text{Linear}_C(X')$ 
    \State $X'' \gets \text{SSM}(A, B, C)(X')$ \Comment{State space modeling}
    \State $H \gets \text{Linear}_T(X'') + X$ \Comment{Residual connection}
\EndFor 
\State \textbf{Cross-Podal Fusion:}
\State $H_S \gets \text{Linear}_{\text{align}}(H_{\theta_S})$\quad $H_T \gets \text{Linear}_{\text{align}}(H_{\theta_T})$ 

\State $H \gets H_S + H_T$

\State \Return $H$
\end{algorithmic}
\end{algorithm}
\vspace{-3mm}
\subsection{Self-supervised ID Classification}

Following the method in \cite{liu2022anomalous}, we adopt a self-supervised ID classification strategy that combines the ESTgram with metadata (i.e., machine types and machine IDs). The dual-modal fusion features $H\in\mathbb{R}^D$ extracted by the Dual-Path Mamba Architecture are subsequently fed into a fully connected layer with a softmax activation function to generate class probabilities. Instead of the traditional cross-entropy loss, we use the ArcFace loss \cite{deng2019arcface} to increase inter-class separability and tighten intra-class distributions. The negative log probability of each audio sample is ultimately used as the anomaly score.

\vspace{-3mm}
\section{Experiment and Result}
\label{sec:Experiment and Result}

\subsection{Dataset}

We conducted experiments on the DCASE 2020 Task 2 development and additional datasets, which include the MIMII \cite{Purohit_DCASE2019_01} and ToyADMOS \cite{Koizumi_WASPAA2019_01} datasets. The dataset contains six types of machines. Except for ToyConveyor, which includes six different machine IDs, all other machine types have seven different machine IDs. In total, there are 41 distinct machine IDs, with each audio signal being a single-channel recording approximately 10 seconds long.

The training set, consisting of normal sound recordings from the development and additional datasets, was used for training, while the test data (including both normal and anomalous sounds) from the development dataset was used for evaluation. We did not use the latest version of the DCASE datasets in our experiments because they primarily focus on domain adaptation and generalization studies.

\vspace{-3mm}
\subsection{Experimental Setting}

\textbf{Implementation Details:}
The input spectrograms are obtained using Short-Time Fourier Transform (STFT) with a window size of 1024 and a hop length of 512. The number of Mel filters is set to 128. In the TSG module, the gating parameter scale is set to 2. For patch segmentation, the number of time patches $I$ is set to 12, and the number of frequency patches $J$ is set to 16. The activation function used is SiLU \cite{elfwing2018sigmoid}. The network is trained using the AdamW optimizer \cite{Loshchilov2017DecoupledWD} with a learning rate of 0.0001 for 200 epochs and a batch size of 128.\\
\textbf{Evaluation Metrics:}
For the evaluation metrics, we followed the standards defined in \cite{Koizumi_DCASE2020_01,suefusa2020anomalous,dohi2021flow} and used the Area Under the Receiver Operating Characteristic Curve (AUC) and the partial AUC (pAUC). The pAUC represents the AUC value within the range of low false-positive rates $[0,p]$, where $p$ is set to 0.1. Additionally, we compared the arithmetic mean of the AUC and pAUC across all machine types to evaluate the method's stability.

\begin{table*}[h]
\centering
\caption{Performance comparison of AUC (\%) with different input features and modules.}
\resizebox{0.75\linewidth}{!}{
    \begin{tabular}{c|c|cccccc|c}
        \toprule
        Methods & Input Feature &  Fan & Pump & Slider & Valve & ToyCar & ToyConveyor & Average \\
        \midrule
        \multirow{2}{*}{SMamba} 
        & STgram & 92.36 & 89.51 & 98.02 & 96.20 & 92.95& 80.28 & 91.55 \\
        & ESTgram & 93.43 & 93.29 & 98.68 & 98.74 & 95.96 & \textbf{82.79} & 93.82 \\
        \midrule
        \multirow{2}{*}{TMamba} 
        & STgram & 95.76 & 90.68 & 97.12 & 98.10 & 94.75 & 75.82 & 92.04 \\
        & ESTgram & 97.36 & 93.88 & 99.40 & 98.94 & 95.56 & 78.43 & 93.94 \\
        \midrule
        \multirow{2}{*}{STMamba} 
        & STgram & 98.30 & 94.90 & 99.26 & 99.52 & 96.69 & 80.75 & 94.90 \\
        & ESTgram & \textbf{98.85} & \textbf{96.39} & \textbf{99.74} & \textbf{99.92} & \textbf{97.28} & 82.35 & \textbf{95.76} \\
        \bottomrule
    \end{tabular}
}
\label{tab:tab3}
\end{table*}

\vspace{-0.3cm}
\subsection{Performance Comparison}

To demonstrate the performance of the proposed ESTM, we compared it with state-of-the-art methods on the DCASE 2020 Task 2 dataset, including MobileNetV2 \cite{giri2020self}, Glow\_Aff \cite{dohi2021flow}, STgram-MFN  \cite{liu2022anomalous}, CLP-SCF \cite{zhang2024dual} , ASD-AFPA \cite{zhang23fa_interspeech} and TASTgram(NAMix) \cite{Noisy-Arcmix}. The results for the other methods in Table 1 are the best reported results from their respective papers. As shown in Table I, the proposed ESTM outperforms previous methods in most machine types, especially on ToyConveyor, where the AUC increased by 7.99\% and pAUC increased by 4.50\%. Furthermore, ESTM achieved the best average AUC and pAUC across all machine types, with an AUC improvement of 0.84\% and a pAUC improvement of 1.79\% over the state-of-the-art ASD-AFPA method.

\vspace{-0.2cm}
\begin{figure}[H]
    \centering
    \begin{subfigure}[b]{0.3\textwidth}

        \centering
\setlength{\belowcaptionskip}{-1pt}

        \includegraphics[width=\textwidth]{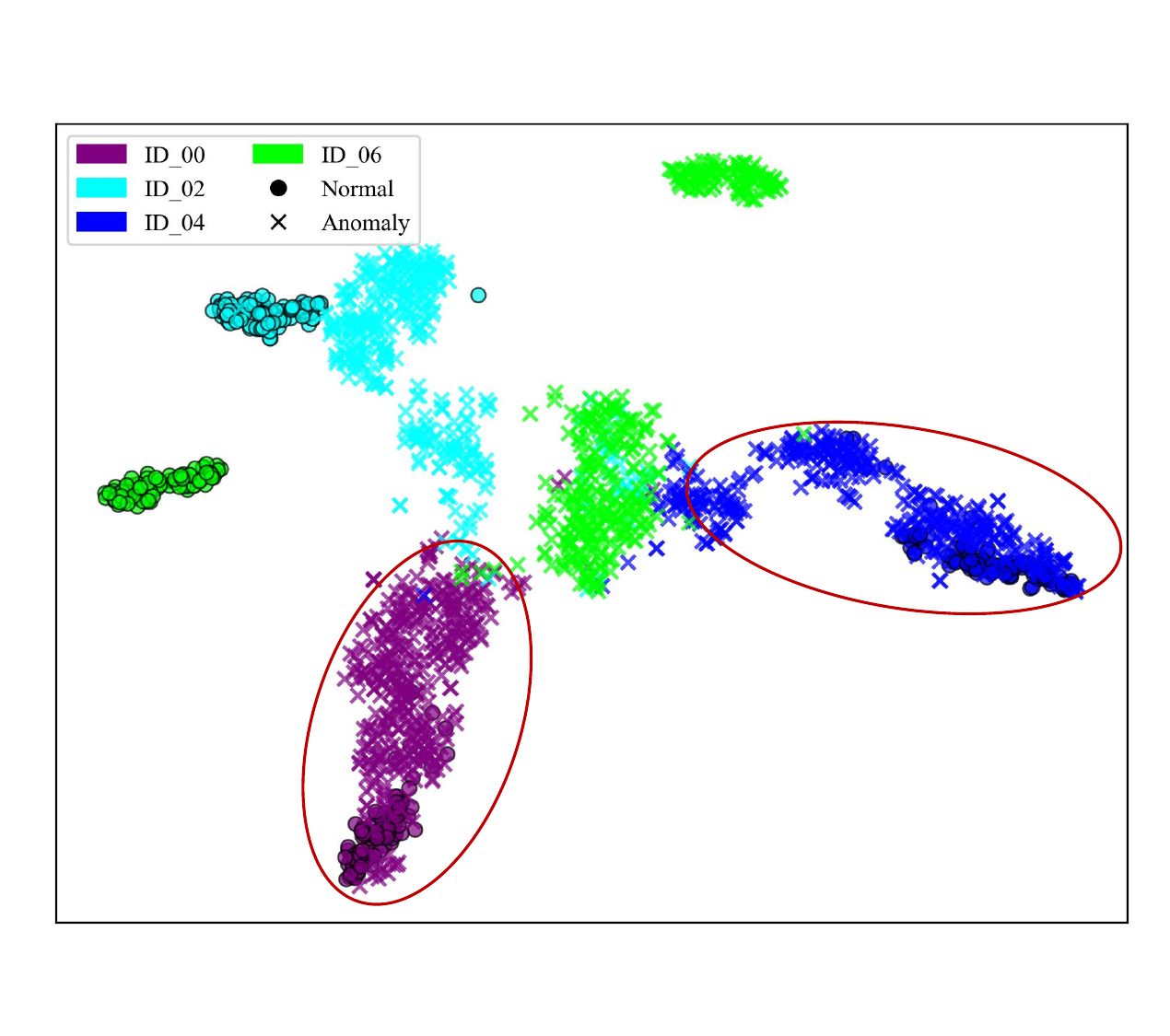} 
        \caption{STMamba (STgram)}
        \label{fig:stgram-mfn}
    \end{subfigure}
    \begin{subfigure}[b]{0.3\textwidth}
        \centering
        \includegraphics[width=\textwidth]{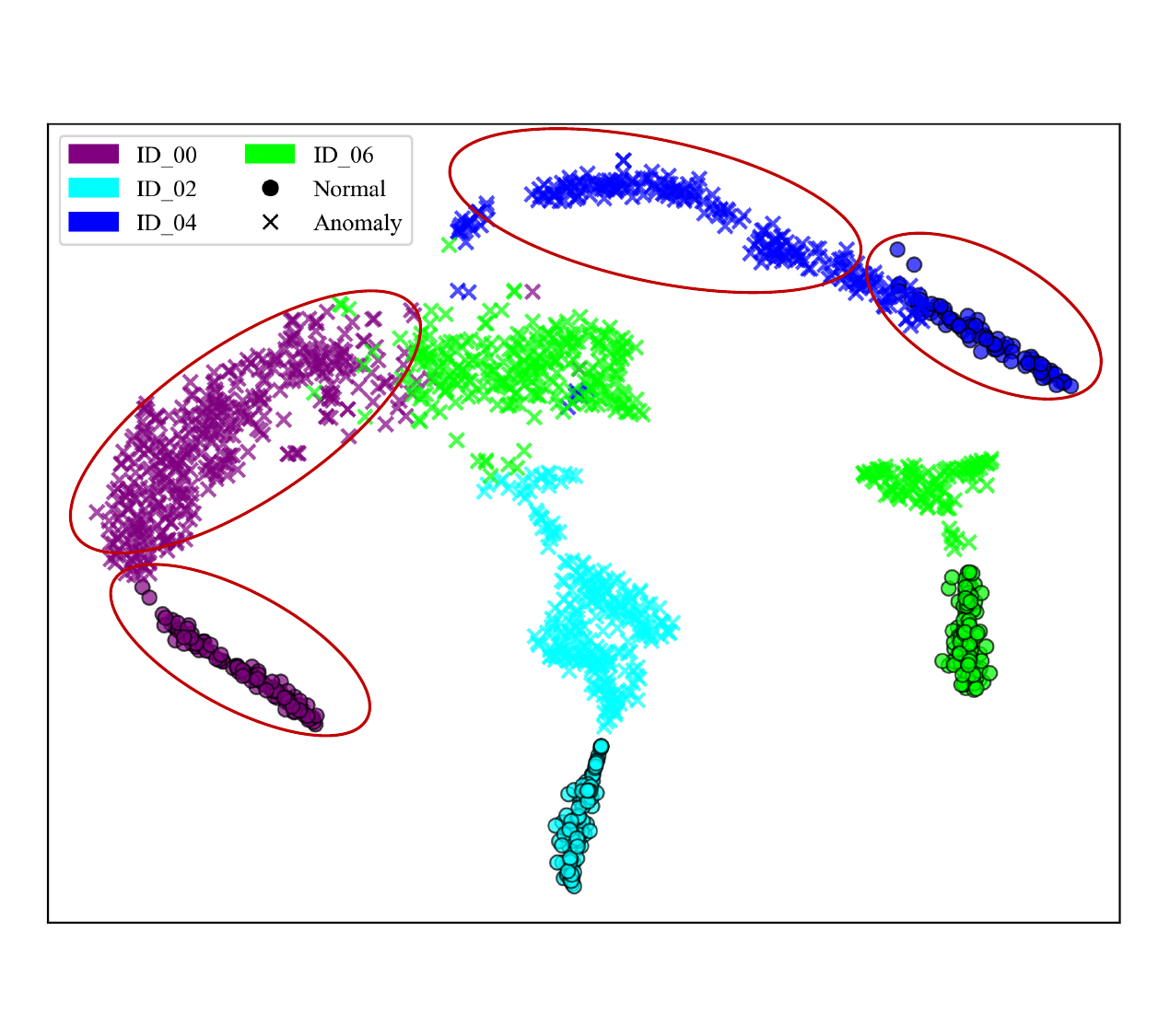} 
        \caption{STMamba(ESTgram)}
        \label{fig:asd-afpa}
    \end{subfigure}
    \vspace{-1mm}
    \caption{The t-SNE visualization comparison of STMamba under different input conditions for the Fan machine type. (a) The latent feature distribution obtained using STgram. (b) The latent feature distribution obtained using ESTgram. The symbols “•” and “×” denote normal and anomalous sound classes, respectively.}
    \label{fig:tsne-comparison}
\end{figure}

\vspace{-3mm}
\subsection{Ablation Study on Feature Representation Stability}

To validate the synergistic effect of the feature enhancement module and the dual-path architecture in the ESTM framework, we conducted a hierarchical ablation experiment. We compared the detection performance of traditional time-frequency features (STgram) and TSG-enhanced features (ESTgram) within SMamba, TMamba, and the spectral-temporal joint modeling branch (STMamba).  As shown in Table II, ESTgram as input consistently improved detection across all machine types, with particularly significant gains for ToyConveyor. For STMamba, the average AUC across all machines increased by 0.91\%, verifying that the TSG module effectively optimizes time-frequency features, generating more discriminative spectral-temporal representations. Moreover, under the same input conditions, STMamba achieved an additional 2.07\% and 1.94\% increase in average AUC compared to SMamba and TMamba, respectively. This demonstrates that our time-frequency decoupled modeling approach successfully captures long-period operational patterns and cross-band energy transfer, effectively achieving complementary feature fusion. This anomaly on ToyConveyor is attributed to its spectrally-dominant anomaly patterns, where temporal fusion paradoxically dilutes critical frequency-domain features.

To visually compare the impact of input representations on the latent feature space, we applied t-distributed stochastic neighbor embedding \cite{cieslak2020t} (t-SNE) to perform dimensionality reduction and visualization of the latent features generated by STgram and ESTgram. As shown in Fig. 2, compared to STgram (a), ESTgram (b) exhibits more pronounced feature separability in the clustering distribution of the Fan machine type. The latent features of normal and anomalous samples in the ESTgram space present clearer inter-class boundaries, particularly in the ID\_00 and ID\_04 clustering regions, where the overlap between anomalous and normal samples is further reduced. This further validates the effectiveness of our proposed method.

\vspace{-2mm}
\section{Conclusion}
In this paper, we propose a novel anomalous sound detection framework, ESTM, which is based on a dual-path Mamba architecture with time-frequency decoupled modeling. This framework enables more fine-grained extraction of machine acoustic features across different frequency bands and time segments. Additionally, the proposed TriStat-Gating (TSG) module enhances the model’s sensitivity to anomalous patterns. Experimental results demonstrate that the proposed method achieves better detection performance than state-of-the-art approaches on the DCASE 2020 Task 2 dataset, further validating the effectiveness of our method.

\newpage

\bibliographystyle{IEEEbib}
\bibliography{mybib.bib}

\end{document}